\documentclass{ar2e}
\usepackage{latexsym,graphicx,epsfig,psfrag,here,color,citesort,amsmath}
\usepackage[comma,sort&compress]{natbib}
\usepackage{citesort}
\newcommand{\eq}{\begin{eqnarray}}
\newcommand{\en}{\end{eqnarray}}
\renewcommand{\theequation}{\arabic{section}.\arabic{equation}}

\newcommand{\fs}{\, .}
\newcommand{\scs}{\, , \,}
\newcommand{\bea}{\begin{eqnarray}}
\newcommand{\eea}{\end{eqnarray}}
\newcommand{\ed}{\end{document}}

\newcommand{\kad}{$\bar K d$\,\,}

\begin{document}
\thispagestyle{empty}

\jname{Ann. Rev. Nucl. Part. Sci.}
\jvol{vol. {59}}
\jyear{2009}

\title{
\begin{flushright}
{\large\rm Preprint HISKP-TH-08/20} 
\end{flushright} 
Hadronic Atoms
}

\markboth{Gasser, Lyubovitskij,  Rusetsky}{Hadronic Atoms}

\author{\hspace{1cm}J.~Gasser$^a$, V.E.~Lyubovitskij$^{b,\dagger}$ and 
A. Rusetsky $^{c}$
\affiliation{
\hspace{1.5cm}\footnotesize{\begin{tabular}{c}
$^a\,$Center for Research and Education in Fundamental Physics,\\[-2mm]
  Institute for Theoretical Physics,  University of Bern,\\[-2mm]
Sidlerstr. 5, CH-3012 Bern, Switzerland\\[2mm]
$^b\,$
Institut f\"ur Theoretische Physik, Universit\"at T\"ubingen,\\[-2mm]
Kepler Center for Astro and Particle Physics, \\[-2mm]
$\hspace{2mm}$Auf der Morgenstelle 14, D72076 T\"ubingen, Germany\\[2mm]
$^c\,$ 
Helmholtz--Institut f\"ur Strahlen-- und Kernphysik,\\[-2mm]
Bethe Center for Theoretical Physics,\\[-2mm]
Universit\"at Bonn,  D--53115 Bonn, Germany\\[2mm]
$^\dagger\,$
On leave of absence from: Department of Physics, \\[-2mm]
$\hspace{2mm}$ Tomsk State University, 634050 Tomsk, Russia
\end{tabular}  }}}

\begin{keywords} 
QCD, QED, chiral perturbation theory,  DGBT formula,\\[-2mm] 
non-relativistic effective Lagrangians, scattering lengths 
\end{keywords}

\begin{abstract}

\baselineskip 12pt 
We review the theory of hadronic atoms in QCD+QED.
The  non-relativistic effective Lagrangian approach,
used to describe this type of bound states, is illustrated
with the case of   $\pi^+\pi^-$ atoms.
In addition, we  discuss the evaluation of  isospin-breaking corrections to  hadronic
atom observables by invoking chiral perturbation theory.

\vspace*{1.9cm}

\noindent\footnotesize{
\begin{center}Commissioned article for {\it Ann. Rev. Nucl. Part. Sci}
\end{center}}
\end{abstract}

\maketitle 

\topmargin 2cm

\baselineskip 18pt 

\newpage 
\renewcommand{\theequation}{\arabic{section}.\arabic{equation}}
\setcounter{equation}{0}

\section{Introduction}

Hadronic atoms are bound states of hadrons, held together 
predominately by the static Coulomb force.  Simple examples 
 are  {\em pionium}, a bound state of two pions with opposite electric charge ($\pi^+,\pi^-$),
 and {\em pionic hydrogen}, 
a bound state of  a $\pi^-$  and a proton. Pionium 
is the analogue of positronium in Quantum Electrodynamics. 
A more complex example is 
 {\em pionic deuterium} -- a Coulombic bound state of a $\pi^-$
and a deuteron. The latter itself is  a composite state of a proton and a 
neutron, bound at much smaller distances than the size of the hadronic atom.

The average distance between the constituents of a hadronic atom  is given by
the Bohr radius
\eq\label{eq:Bohr}
r_B=\frac{1}{\alpha\mu_c}\, ,\quad\quad \mu_c=\frac{M_1M_2}{M_1+M_2}\, ,
\en
where $\alpha\simeq 1/137$ is the fine-structure constant, $\mu_c$ stands for 
the reduced
mass, and $M_1,M_2$ are the masses of the  constituents.
Typically, the Bohr radius is of the order of a few hundred Fermi, much larger than
the range of strong interactions. Individual hadrons in the atom
spend most of the time at  distances where  strong interactions are practically
absent. For this reason, observables of hadronic atoms are barely affected by 
the strong interactions.

\begin{figure}[t!]
\begin{center}
\includegraphics[width=5.cm]{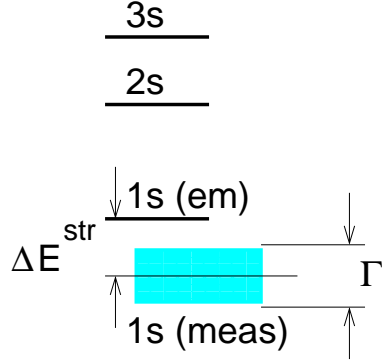}
\end{center}
\caption{
\baselineskip 12pt 
Schematic representation 
of the shift of the ground state energy level in  pionium.
$\Delta E^{\rm str}$ and $\Gamma$ denote the strong energy shift and the 
width in the ground state (we omit indices in $\Gamma$ throughout, because we will  
consider the widths of ground states only). The symbols {\em em} and 
{\em meas} denote the purely electromagnetic and the measured energy levels, respectively.}
\label{figure1}
\end{figure}

If the interaction between the constituents were purely electromagnetic 
and non-relativistic,
the energy levels of the atom would be given by the standard quantum-mechanical
formula
\eq
E_n=M_1+M_2-\frac{1}{2n^2}\,\mu_c\alpha^2\, ,\quad\quad n=1,2,\cdots\, \fs
\en
 Aside from relativistic corrections which generate higher order terms in $\alpha$, 
this formula is modified in the presence of strong interactions in two ways.
 First, the energy levels are shifted from their purely electromagnetic value.
Furthermore, because the atoms can  decay also via strong interactions
(example: the decay
of  pionium into a neutral pion pair through the charge-exchange reaction
$\pi^+\pi^-\to \pi^0\pi^0$),  the energy levels are broadened.
The effect on the ground-state energy level is illustrated in 
Fig.~\ref{figure1}. 
The pionium lifetime in the ground state, 
$\tau=1/\Gamma\simeq 3\times 10^{-15}~{\rm s}$, is still
much smaller than the charged pion lifetime $\tau_\pi\simeq 10^{-8}~{\rm s}$. 
 Despite the short lifetime of the atom, 
the pions travel many times around each other 
before the atom decays, as the ratio $\frac{1}{2}\mu_c\alpha^2/\Gamma\simeq 8\times 10^3$ 
indicates. 
As a consequence of this, pionium can be considered  a 
quasi-stable bound state with a clearly defined structure of  
(almost Coulombic) energy levels. The same statement is valid for many other
hadronic atoms.

Because the size of hadronic atoms is much larger than
the range of strong interactions, the energy levels of the atoms can depend
only on the characteristics of hadronic interactions at asymptotically
large distances. These are usually described in terms 
of the parameters in the effective range expansion: scattering
length, effective range, shape parameters. 
The situation is analogous to the calculation of the classical 
static electric field generated by a charge distribution:
 at asymptotic distances, the electric field
depends only on the multipole moments which describe the charge
distribution.

Deser, Goldberger, Baumann and Thirring (DGBT) were the first to derive
 -- at leading order in the fine-structure constant $\alpha$ -- a formula 
for the complex shift of the energy level of a hadronic atom~\cite{Deser}.
The real and imaginary parts of this shift
 define the displacement and the width of
a given level, generated by the  strong interactions. 
The formula for the ground state reads
\eq\label{eq:Deser}
\Delta E^{\rm str}-\frac{i}{2}\, \Gamma=-2\alpha^3\mu_c^2\, 
{\cal T}+\cdots\, ,
\en
where ${\cal T}$ denotes the complex elastic scattering amplitude
of the constituents at  the threshold.
The ellipses stand for higher order isospin breaking corrections, 
which will be discussed in detail later in this article.
 The formula can be trivially generalized to the case of excited energy levels.

The DGBT formula~(\ref{eq:Deser}) plays a central role in the theory of
hadronic atoms, because it  allows one to extract 
the threshold amplitude ${\cal T}$ from the experimentally measured
energy and width of the atom.
Further, the real and imaginary parts of 
${\cal T}$ are  related to
 the  hadronic scattering lengths.
From this we conclude that the experimental study of hadronic atoms 
provides us with a source for the determination of
these scattering lengths.
 The above-mentioned
 huge difference in the atomic and strong interaction scales is very
advantageous in the present context, since the atomic observables depend 
(at leading order) exactly
 on those quantities  (scattering lengths)
which one wants to extract from the experiment -- they
 are not sensitive to the short-range details of strong interactions.

In most 
cases, the first term on the right-hand side of  Eq.~\eqref{eq:Deser}
 does not match the available experimental  precision -- next-to-leading 
order corrections, 
indicated with the ellipses in Eq.~\eqref{eq:Deser}, are needed as well for a precise determination of the scattering lengths. The aim of
any theory of hadronic atoms must be to provide a {\em systematic} framework for the
calculation of these corrections. Here, we will carry out the 
calculations  by using a  non-relativistic 
effective theory of Quantum Chromodynamics + Quantum Electrodynamics 
(QCD+QED).
 We will illustrate the method  by means of  pionium decay,
and will briefly consider the application of the same approach to 
pionic hydrogen and pionic deuterium.
 We no not discuss  quantum-mechanical potential models, 
because   these methods introduce inherent model-dependent 
artefacts which cannot be  controlled.

The article is organized as follows. In section~\ref{sec:background}
 we consider the physics background
behind the experiments on various hadronic atoms. 
In section~\ref{sec:essentials} we briefly discuss the essentials of the 
non-relativistic effective Lagrangian approach, which is  our
tool to describe hadronic atoms. Section~\ref{sec:pionium} forms
the backbone of the present article. In this section we construct, step by step,
the effective field theory approach to  pionium decays. The same approach
is applied in section~\ref{sec:piH} to the description of pionic hydrogen
and of pionic deuterium. Finally, section~\ref{sec:concl} contains a brief summary
and outlook for  future research in the field.

\setcounter{equation}{0}
\section{Physics background}
\label{sec:background}

Several experiments with hadronic atoms are presently running.
 The DIRAC collaboration at CERN is measuring
 the lifetime of
 pionium~\cite{Trautmann1,DIRAC1,Trautmann2,DIRAC2,DIRAC3,DIRAC-2.9,Zhabitsky}  
and plans to determine the lifetime of  $\pi K$
atoms as well~\cite{DIRAC-piK}.  The Pionic Hydrogen collaboration 
at PSI~\cite{PSI1,PSI5,PSI2,PSI3,PSI4}
 studies the spectrum of pionic hydrogen and  pionic deuterium,
whereas the DEAR/SIDDHARTA collaboration 
at LNF-INFN~\cite{SIDDHARTA1,SIDDHARTA2,SIDDHARTA4,SIDDHARTA3} 
 plans to determine the ground state energy and  width
 of  kaonic hydrogen at a much better accuracy than in  previous experiments
 carried out at KEK~\cite{KEK1,KEK2}. In addition,  SIDDHARTA plans 
the first ever measurement of the  spectrum of kaonic deuterium.

These experiments eventually result in a precise determination of
various hadro\-nic scattering lengths.
 Let us recall why the results will be important  for the investigation of 
 several fundamental properties of QCD.

1. We start   with the DIRAC experiment 
at CERN~\cite{DIRAC1,DIRAC2,DIRAC3,DIRAC-2.9,Zhabitsky,Trautmann1,Trautmann2}. 
The decay width of  the ground state of pionium  into a $\pi^0\pi^0$ pair 
is related to the difference of
the $S$-wave $\pi\pi$ scattering lengths $a_0,a_2$
with  total isospin 0 and 2,
\eq\label{eq:pionium}
\Gamma=\frac{2}{9}\, \alpha^3 p^\star(a_0-a_2)^2+\cdots\fs
\en
Here, $p^\star=(M_\pi^2-M_{\pi^0}^2-\frac{1}{4}\, M_\pi^2\alpha^2)^{1/2}+\cdots$
is the CM momentum of the neutral pion pair after decay,
$M_\pi,M_{\pi^0}$ are the charged and neutral pion masses, respectively,
and the ellipses stand for  terms
of higher order in  isospin breaking.

 It is expected that the DIRAC experiment will finally provide  a value for
$|a_0-a_2|$ which is accurate up to a few percent. Other experiments, where
the $\pi\pi$ scattering lengths are determined from $K_{e4}$ 
decays~\cite{Ke4_1,Ke4_2,Ke4_3,ke4-we,ke4xxx}
or from studying the cusp structures in $K\to 3\pi$ 
decays~\cite{Budini,Cabibbo1,Cabibbo2,K3pi,Gamiz,cuspwe1,cuspwe2,%
ktev,cuspwe3,Kampf}, 
yield competitive results in terms of accuracy.
 On the other hand, the difference $a_0-a_2$ is particularly sensitive  to the
value of the quark condensate in QCD~\cite{small1,small2,small3,small4}. 
In the so-called ``standard'' scenario
which assumes a large condensate,  the expansion of the pion mass in 
terms of the quark mass is
\eq
M_\pi^2=M^2-\frac{\bar l_3}{32\pi^2F^2}\, M^4+O(M^6)\, ,\quad
M^2=2\hat m B\, ,\quad
\hat m=\frac{1}{2}\,(m_u+m_d)\, ,
\en 
where the second term on the right hand side of the first equation 
is small \cite{Colangelo:2001sp}. 
Here, $F$ is the pion
decay constant $F_\pi$ in the chiral limit, $m_u,m_d$ are the light quark 
masses, $\bar l_3$ denotes one of the low-energy constants (LECs) 
in chiral perturbation theory (ChPT, see, e.g., References~\cite{Weinberg_ChPT,GL_ChPT}), and the quantity $B$ 
is related to the quark condensate in the chiral 
limit~\cite{Colangelo:2001sp,GL_ChPT}. 
Further, if  chiral symmetry breaking
in QCD proceeds according to the standard picture, a very accurate description
 of the scattering lengths
 $a_0,a_2$ can be achieved by combining 2-loop 
ChPT with the Roy equations~\cite{Colangelo_scatt1,Colangelo_scatt2},
\eq\label{eq:a0a2}
a_0=0.220\pm 0.005\scs
a_2=-0.0444\pm 0.0010\scs
a_0-a_2=0.265\pm 0.004\fs
\en
Equipped with this precise theoretical prediction, one may perform a 
direct experimental test of the chiral symmetry breaking scenario in QCD. 
Namely, if the measured value of $a_0-a_2$ differs significantly from
the theoretical prediction, this would suggest that 
chiral symmetry breaking proceeds in a manner which is different from the 
standard scenario.  The present situation concerning the verification of  
 the predictions (\ref{eq:a0a2}) is the following. 
Lattice results for $a_2$~\cite{lattice_pipi2,%
lattice_pipi4} agree with the prediction within one standard deviation, see Ref.~\cite{lattice_pipi4} for a compilation of predictions, lattice calculations and data.
 Due to technical difficulties (disconnected graphs),  $a_0$ has
not yet been measured with this technique.
 On the other hand, using the LECs $\bar l_{3,4}$ (or their $SU(3)\times SU(3)$ counterparts) determined 
from the lattice and converting these into a value of $a_0$ again leads to agreement
 with the prediction \eqref{eq:a0a2}, within one standard deviation~\cite{l3l42,%
leut_formula,l3l43,l3l41,l3l46,l3l44,l3l45}. 
We refer the interested reader to Ref.~\cite{leut_formula} for discussions
 and for a review. 
On the experimental side, 
data of the DIRAC collaboration on pionium lifetime~\cite{DIRAC-2.9}, of NA48/2 
on the cusp in $K\to 3\pi$ 
 decays~\cite{K3pi} and  on $K_{e4}$
events~\cite{Ke4_3} (applying  isospin-breaking corrections as described in 
Refs.~\cite{ke4-we,ke4xxx})
 neatly confirm the predictions, 
although partly still with considerable uncertainties.

2. Next, we briefly consider the proposed measurement of the $\pi K$ atom 
lifetime~\cite{DIRAC-piK}, which enables one to extract the value of the
 isospin-odd $S$-wave $\pi K$ scattering length $a_0^-$. The calculations of 
this scattering length, carried out in ChPT up to two loops, lead to a rather 
contradictory picture: it turns out~\cite{piK-2loop1,piK-2loop2} 
 that the two-loop contribution to this quantity
is apparently larger than the one-loop correction. On the other hand,
 the result at two loops agrees with
the analysis carried out on the basis of Roy equations~\cite{piK-Roy}. The situation 
is puzzling, because, if correct,  the convergence of the ChPT series 
for pion-kaon scattering is under question.  It is clear that
a precise knowledge of the  experimental value of the scattering length 
is an important ingredient to the solution of this puzzle. For more comments 
concerning this point, we refer the interested reader to section 2 of the 
review~\cite{Physrep}.

3. From the measurement of the pionic hydrogen energy shift and width by 
the Pionic Hydrogen collaboration at PSI~\cite{PSI1,PSI2,PSI3,PSI4,PSI5}, one 
can extract the isospin even and odd $S$-wave $\pi N$ scattering lengths
$a_{0+}^+$ and $a_{0+}^-$. 
Using Effective Field Theory methods (EFT) in the two-nucleon sector, one
can also relate the pion-deuteron scattering lengths to the pion-nucleon ones.
 [See, e.g., Refs.~\cite{Weinberg-deuteron,Bernard,Borasoy-Beane2,Borasoy-Beane1,%
Bernard2_1,Hanhart_Ko2,Doring,Raha2,Raha3,Epelbaum,Valderrama-Platter2,%
Valderrama-Platter1,Bernard2_2,Hanhart_Ko1,Bernard2_3}. 
A similar result can be obtained
with quantum-mecha\-nical multiple-scattering 
theory (see, e.g., Refs.~\cite{Ericson-Weise2,Ericson-Weise1,Ericson-Weise3}).]
Thus, the measurement of the energy shift and width of pionic deuterium
results in  additional constraints
on the values of $a_{0+}^+$ and $a_{0+}^-$. 

 $\pi N$ scattering lengths
are quantities of fundamental importance in low-energy hadronic physics 
by themselves,
since they test the exact pattern of explicit chiral 
symmetry breaking. Moreover,  knowledge of the exact values of the 
scattering lengths also affects our understanding 
of more complicated systems where the $\pi N$ interaction serves
as an input, e.g. $NN$ interaction, pion-nucleus scattering, three-nucleon
forces, etc.
In addition,  high-precision values of the $\pi N$ scattering 
lengths are used as an input for the determination of different basic
parameters of QCD at low energies more accurately. One example 
is the $\pi NN$ coupling constant $g_{\pi NN}$, which is 
obtained from the Goldberger-Myazawa-Oehme (GMO) 
sum rule~\cite{GMO,GMO-recent1,GMO-recent2}, where
a particular combination of scattering lengths enters as a subtraction
constant.
Other important quantities, which can be obtained by using 
the $S$-wave $\pi N$ scattering lengths as an input, are the so-called pion-nucleon
sigma-term and the strangeness content of the nucleon. 
The sigma-term $\sigma_{\pi N}$,
which measures the explicit breaking of chiral symmetry in the one-nucleon
sector, is defined by
\eq\label{eq:scalar}
\sigma_{\pi N}=\frac{1}{2m_N}\langle ps|\hat m(\bar uu+\bar dd)|ps\rangle
\, ,
\en
where $|ps\rangle$ denotes a one-nucleon state, with momentum $p$ and  spin $s$, 
and $m_N$ is the nucleon mass.
 The sigma-term is related to the strangeness content $y$
of the nucleon, and to the $SU(3)$ symmetry breaking part of the strong
Hamiltonian,
\eq\label{eq:y}
&&\frac{m_s-\hat m}{2m_N}\,\langle ps|\bar uu+\bar dd-2\bar ss|ps\rangle=
\biggl(\frac{m_s}{\hat m}-1\biggr)(1-y)\sigma_{\pi N}\, ,
\nonumber\\[2mm]
&&y=\frac{2\langle ps|\bar ss|ps\rangle}{\langle ps|\bar uu+\bar dd|ps\rangle}\, ,
\en
where $m_s$ denotes  the strange quark mass.
In the analysis of the ex\-peri\-men\-tal data, one uses
 $S$-wave $\pi N$ scattering lengths as 
input in the dispersion relations, which provide the extrapolation of the
isospin even pion-nucleon scattering amplitude from threshold down to the
Cheng-Dashen point. We refer the interested reader to Ref.~\cite{sigma45} for 
details. In this reference, the value $\sigma_{\pi N}\simeq 45$ MeV was obtained. 
[In Ref.~\cite{pavan}, a value for the sigma-term which is considerably larger 
than 45 MeV  is claimed to follow from more recent data.] 
 The sigma-term  is rather sensitive to the scattering lengths~\cite{sigma45}.
 Consequently, an accurate measurement of the latter 
 will have a large impact on the experimentally determined 
values of $\sigma_{\pi N}$ and $y$.
 Finally, we note that the sigma-term is accessible through lattice calculations, 
see e.g. Ref.~\cite{sigmalattice} and references cited there. It even 
 plays a role in astrophysical applications.
As an example for such an impact,  we refer the interested reader 
to the recent publication~\cite{astrophysical} and the references given there.

4. Last but not least, we discuss the DEAR/SIDDHARTA experiment 
at LNF-INFN~\cite{SIDDHARTA1,SIDDHARTA2,SIDDHARTA3,SIDDHARTA4}.  
It plans to determine $\bar KN$ scattering lengths from data 
on kaonic hydrogen and on kaonic deuterium atoms.
 We believe that
it would be very useful to carry out a comparison of the scattering lengths 
 so determined with different theoretical predictions based on the unitarization
of the lowest order ChPT amplitude
\cite{Siegel1,Siegel2,MO,Siegel3,Raha1,Nissler1,Nissler2,OPV,BMN}.
 Indeed, it turns out that even the data from kaonic hydrogen alone
 impose rather stringent 
constraints on the values of the $\bar KN$ scattering lengths.  
In some cases, 
DEAR/SIDDHARTA data seem not to be compatible with the
scattering sector~\cite{Raha1,Nissler1,BMN}.
It is clear that imposing additional constraints from \kad data
makes the issue  even more pronounced. 
In our opinion, it is  important to check 
whether the unitarization approach passes this test.

 \setcounter{equation}{0}

\section{The non-relativistic effective theory}
\label{sec:essentials}

At leading order, the DGBT formula in Eq.~(\ref{eq:Deser}) is universal: it looks exactly
the same  in  potential scattering theory (where it was derived first) and
in quantum field theory. This fact is due to the   huge difference between the
atomic and the strong interaction scales  mentioned in the
introduction.  On the other hand, the isospin-breaking corrections to this 
relation, which are due to  electromagnetic 
interactions and to the quark mass difference $m_d-m_u$,  are not universal.
 In this article, we describe a
systematic theory of hadronic atoms within QCD+QED, which enables one to
calculate these corrections in a simple and elegant manner, 
with an accuracy that matches the experimental precision. [
Because ChPT is the low-energy effective theory of QCD+QED, 
one might be tempted to start from this framework.
However, describing bound states in ChPT by using standard techniques, 
based on the Bethe-Salpeter
equation or on 3-dimensional reductions  
thereof~\cite{ivanovm1,Sazdjian1,ivanovm2,ivanovm3,Sazdjian2,Sazdjian3},  
is a  complicated enterprise, which makes it very difficult
to reach the required precision. We do not, therefore, discuss this method here.]

Our framework is based on the 
existence of several different momentum scales in the problem.
 Counting powers of the fine-structure constant $\alpha$, we have
to assign the order $\alpha^0$ to the  scale related to the pion mass, 
because $M_\pi$ has a  non vanishing value also in the absence of 
electromagnetic interactions.
  On the other hand, the momentum scale
corresponding to atomic phenomena is given by the inverse Bohr
radius  -- i.e., the average 3-momenta inside the atom are 
$p_{\rm av}\simeq r_B^{-1}=\alpha\mu_c$, and count as order $\alpha$. From this one
 concludes that a {\em non-relativistic}
approach,  based on an expansion
in (small) momenta, is  the appropriate framework to describe hadronic
atoms, because the momentum expansion translates into an expansion in 
the fine-structure constant  for  hadronic atom observables. 
 The advantage of considering a non-relativistic framework consists in the
simple treatment of  bound states: they can be described by 
the Schr\"odinger equation\footnote{Caswell and Lepage~\cite{Caswell} were the first to use a systematics non-relativistic effective Lagrangian approach to investigate
 bound states in QED.}.

Let us list some very general properties of this approach.

\begin{itemize}
\item[i)]
The framework uses the language and methods of (non-relativistic) 
quantum field theory.
In particular, the calculations are based on 
effective Lagrangians and Hamiltonians.
\item[ii)]
The non-relativistic approach allows one to keep 
the number of heavy particles  conserved, by construction. In other words,
one always stays within a restricted sector in Fock space.
\item[iii)]
The non-relativistic theory describes  matrix elements at small external momenta.
All high-energy effects -- like transitions to  sectors with a different number of 
heavy particles -- are encoded in the coupling constants of the
effective  Lagrangian, which are determined through  matching to the underlying theory. 
 In this manner, one makes sure that the effective and the underlying theory
 are equivalent at low energies.

\item[iv)]
{\em Power counting rules} are at the heart of any effective field theory.
The non-relativistic power counting at tree-level amounts to counting 
the number of space derivatives in various terms. Because
 each non-relativistic momentum is of order
of $\alpha\mu_c$, the contributions to the bound-state energy from
 terms containing  higher derivatives are
suppressed by additional powers of $\alpha$. 
To carry out calculations of the bound-state 
energy spectrum at a fixed order in $\alpha$,
a finite number of terms in the Lagrangian thus suffices [for comparison, 
in ChPT the number of the relevant terms is infinite].  
\item[v)]
The non-relativistic Lagrangian is used to generate
Feynman graphs in a standard manner.
Strong loops respect the power counting, if dimensional regularization is
used. Loops with photons can also be made consistent with power counting 
by applying the so-called {\em threshold expansion}~\cite{Beneke1,Beneke2}. 
\item[vi)]
It is useful to extend the power counting to include the isospin breaking
effects which are generated by the quark mass 
difference $m_d-m_u$ as well -- along with the
electromagnetic corrections characterized by the fine-structure constant
$\alpha$. There is no strict rule for doing this. We now note that the effect of $m_d-m_u$ in the pion mass is of  order $(m_d-m_u)^2$, and linear for kaons and nucleons.
It is therefore convenient to
introduce a {\em common} isospin-breaking parameter $\delta$ and
count $\alpha\sim(m_d-m_u)^2\sim\delta$ in pionium,
$\alpha\sim(m_d-m_u)\sim\delta$ otherwise. This has the advantage that the
leading corrections to the hadronic atom observables, generated
by $\alpha$ and $m_d-m_u$, are counted at the same order in $\delta$.
  
\item[vii))]
At the end of the day, when the hadronic atom spectrum is calculated and
the matching to the underlying theory is performed, there is no reference left to the
non-relativistic theory in the final result. The non-relativistic approach is
used only at an intermediate stage, in order to facilitate the calculations. 
\end{itemize}

Our main goal here is to  {\it first}
evaluate the $O(\delta)$ isospin-breaking corrections to the leading order 
strong energy shift and width. These corrections are 
 indicated by the ellipses in the DGBT formula
 Eq.~(\ref{eq:Deser}). Due to lack of space, we  concentrate on those corrections 
that are relevant for the {\it width} of the ground state. These are more easy to pin 
down than those for 
the real part of the energy shift. In a {\it second} step, the right hand side 
of  equation Equation~\eqref{eq:Deser} will be 
expressed in terms of isospin symmetric scattering lengths, up to isospin
 breaking corrections. The path to a comparison of the DGBT formula with 
experimental data is then paved, 
and a precise determination of scattering lengths 
becomes feasible.

In the next section, we consider in some detail 
the construction of a non-relativistic theory along these lines.
 The framework  was developed during the last decade in 
Refs.~\cite{Labelle1,Labelle4,Labelle3,Bern1,Labelle5,Labelle2,%
Bern2,Bern3,Bern4,Mojzis,SchweizerHA1,SchweizerHA2,SchweizerHA3,%
Zemp,Raha1,Raha2,Raha3,Raha4,Physrep}.  

\setcounter{equation}{0}
\section{Pionium: Decay of the ground state}
\label{sec:pionium}

Instead of presenting the non-relativistic effective theory in its full generality, 
we have decided to explain the method with one particular example,
 the decay of  pionium. 
Technical details will be  skipped -- these can be found,
e.g., in Refs.~\cite{Bern1,Bern4,Physrep}. For a thorough
discussion of the 
properties of non-relativistic theories,  we refer the reader 
to Refs.~\cite{Physrep,Antonelli}.

\subsection{Non-relativistic framework: strong sector}

We start with a non-relativistic theory for pions, in the absence of photons, which will 
  be included afterwards.
  On the other hand, it is very convenient to keep from the beginning
 the masses of charged and neutral pions at their
physical values. This is a perfectly consistent procedure, because in
the non-relativistic theory, these masses are not renormalized, even when
the electromagnetic interactions are turned on.

\begin{figure}[t!]
\begin{center}
\includegraphics[width=6.cm]{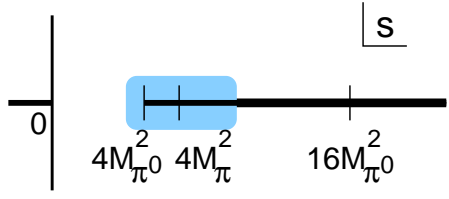}
\end{center}
\caption{\baselineskip 12pt 
The singularity structure of the $\pi\pi$ partial-wave scattering 
amplitudes in the complex $s$-plane. The shaded area
denotes the low-energy domain.}
\label{figure2}
\end{figure}

\begin{sloppypar}

Our starting point is the relativistic amplitude for the process 
$\pi^a(p_1)\pi^b(p_2)\to\pi^c(p_3)\pi^d(p_4)$, where $a,b,c,d=\pm~\mbox{or}~0$.
 Performing a partial-wave expansion, we arrive at the partial-wave 
amplitudes  that
depend on a single Mandelstam variable $s=(p_1+p_2)^2$. 
Assuming further $s$ to be a 
complex variable, we
consider the singularity structure of the partial-wave 
amplitudes in the low-energy region $|s-4M_\pi^2|\ll M_\pi^2$
[For definiteness, we consider here the sector with  total charge $Q=0$.
Other sectors can be discussed analogously.]
As it is well known, the partial-wave  amplitude is holomorphic in the complex $s$-plane, cut along the
positive real axis for  $s \geq 4M_{\pi^0}^2$, see Fig.~\ref{figure2}. 
Another branch point corresponding to 
the two charged pion threshold is located at $s=4M_\pi^2$, whereas   the first inelastic
threshold is located at $s=16M_{\pi^0}^2$. In addition, there is a cut 
on the negative real axis. However, the 
distance between these faraway 
singularities and the 2-pion threshold is of the order
of the pion mass squared. Consequently, in the low-energy region, which includes 
 the neutral and charged two-pion thresholds (far below the 
first inelastic threshold), the 
 partial-wave amplitude has a particularly simple form~\cite{ke4-we},
\eq\label{eq:ABCD}
T_l(s)&=&A_l(s)+iB_l(s)\sigma(s)+iC_l(s)\sigma_0(s)+D_l(s)\sigma(s)\sigma_0(s)\, ,
\nonumber\\[2mm]
\sigma(s)&=&\sqrt{1-\frac{4M_\pi^2}{s}}\, ,\quad\quad
\sigma_0(s)\, \,=\, \, \sqrt{1-\frac{4M_{\pi^0}^2}{s}}\, ,
\en
where
 $A_l(s),\cdots D_l(s)$ are meromorphic functions in the low-energy
domain.
 \end{sloppypar}
We now present a   framework which 
describes the relativistic 
$\pi\pi$ scattering amplitude in the low-energy
region, and thus reproduces this structure of the amplitude.
  The kinetic term in the Lagrangian is fixed through the
non-relativistic expansion of the relativistic one-particle energy,
\eq\label{eq:Lkin}
{\cal L}_{kin}&=&\sum_\pm \Phi_\pm^\dagger\biggl(i\partial_t-M_\pi
+\frac{\triangle}{2M_\pi}+\frac{\triangle^2}{8M_\pi^3}+\cdots\biggr)\Phi_\pm
\nonumber\\[2mm]
&+&\Phi_0^\dagger\biggl(i\partial_t-M_{\pi^0}
+\frac{\triangle}{2M_{\pi^0}}+\frac{\triangle^2}{8M_{\pi^0}^3}+\cdots\biggr)
\Phi_0\, ,
\en
where $\Phi_\pm,\Phi_0$ denote the non-relativistic field operators for the 
charged and for neutral pion fields, respectively. The propagator of the 
non-relativistic charged pion field is given by
\eq
i\langle 0|T\Phi_\pm(x)\Phi_\pm^\dagger(0)|0\rangle
=\int\frac{d^4p}{(2\pi)^4}\,\frac{{\rm e}^{-ipx}}
{M_\pi+{\bf p}^2/2M_\pi-p^0-i0}\, .
\en
The relativistic corrections due to the higher-order terms in
Eq.~(\ref{eq:Lkin}) 
are treated perturbatively. The non-relativistic 
propagator for the neutral pion is obtained by replacing $M_\pi\to M_{\pi^0}$.

\begin{figure}[t!]
\begin{center}
\includegraphics[width=11.cm]{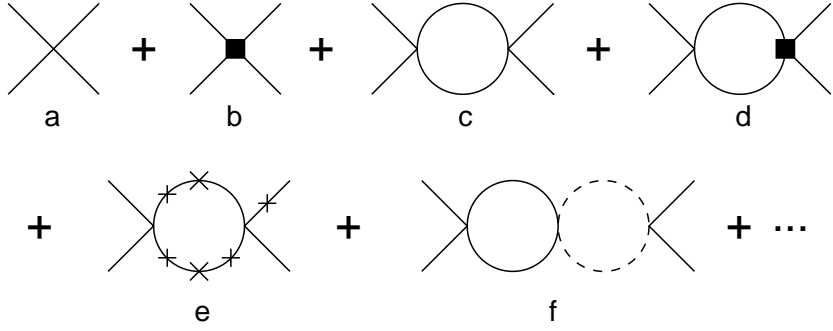}
\end{center}
\caption{\baselineskip 12pt 
Non-relativistic theory. Typical diagrams which contribute to the 
two-pion elastic 
scattering amplitude of the process \eqref{eq:scattering}. 
Solid (dashed) lines: charged (neutral) pions. Filled boxes and crosses denote  derivative
vertices and self-energy insertions, respectively.} 
\label{figure3}
\end{figure}

The free non-relativistic field operators $\Phi_\pm,\Phi_0$ annihilate the 
vacuum. This property can be used to construct a theory that -- from the 
beginning -- conserves the number of pions. 
The interaction Lagrangian  is then given by 
an infinite series of 4-pion local 
operators with an increasing number of space derivatives.
In particular, in the 2-particle sector with  zero total charge -- spanned by the 
states $|\pi^+\pi^-\rangle$ and $|\pi^0\pi^0\rangle$ -- the 
interaction Lagrangian is written as
\eq
{\cal L}_I=c_1\Phi_+^\dagger\Phi_-^\dagger\Phi_+\Phi_-
+c_2(\Phi_+^\dagger\Phi_-^\dagger\Phi_0\Phi_0+\mbox{h.c.})+
c_3\Phi_0^\dagger\Phi_0^\dagger\Phi_0\Phi_0+\cdots\, ,
\en
where the ellipses stand for derivative terms.
The $\pi\pi$ scattering amplitude is calculated 
by using standard Feynman diagram techniques. To be specific, we consider the process
\eq\label{eq:scattering}
\pi^+(p_1)\pi^-(p_2)\rightarrow\pi^+(p_3)\pi^-(p_4)\fs
\en
 Owing to the conservation of the number of pions, 
  the structure of  Feynman diagrams 
is particularly simple: to all orders, the pertinent Green function 
is determined by the bubble diagrams displayed
in Fig.~\ref{figure3}. 
 In the CM frame $P^\mu=p_1^\mu+p_2^\mu=(P^0,{\bf 0})$ ,
the contribution from Fig.~\ref{figure3}c is proportional to
 the loop integral
\eq\label{eq:Jbubble}
J(P^0)&=&\int\frac{d^4l}{i(2\pi)^4}\, 
\frac{1}{M_\pi+{\bf l}^2/2M_\pi-P^0+l^0-i0}\,
\frac{1}{M_\pi+{\bf l}^2/2M_\pi-l^0-i0}
\nonumber\\[2mm]
&=&\frac{iM_\pi p_c}{4\pi}\, ;
\quad p_c=\sqrt{M_{\pi}(P_0-2M_{\pi}) }\fs
\en
We have used  dimensional regularization in  intermediate steps of the calculation 
to tame ultraviolet divergences. Neutral pion loops are obtained with the replacement $M_\pi\rightarrow M_{\pi^0}$.
The contribution to the scattering amplitude for the process 
Eq.~\eqref{eq:scattering} is obtained by putting 
$P_0=2\sqrt{M_\pi^2+\bf{p}^2}=\sqrt{s}$, 
where $\bf{p}$ denotes the pion three momentum in the CM frame.
Therefore, the bubble graphs in Fig.~\ref{figure3} generate polynomials in the quantities
\eq\label{eq:pcp0}
p_c=M_\pi\,\sigma\,[1+O(\sigma^2)]\,,\,\quad 
p_0=M_{\pi^0}\,\sigma_0\,[1+O(\sigma_0^2)]\fs
\en
The so constructed non-relativistic scattering amplitude
 reproduces the 
general low-energy structure of the relativistic amplitude in Eq.~(\ref{eq:ABCD}). 
The counterparts of the functions $A_l(s),\cdots, D_l(s)$  are given 
in form of a power series in $p_c^2,p_0^2$, with  coefficients that depend on 
the non-relativistic couplings $c_1,c_2,c_3,\cdots$.

In order to ensure that the relativistic and the non-relativistic
theories describe the same physics at low energies, it remains to 
 {\em match} the two theories, or, what is the same, to fix the non-relativistic
coupling constants $c_1,c_2,c_3,\cdots$. The matching condition is formulated
for the  $T$-matrix elements,
\eq\label{eq:matching}
T_R^{ab;cd}(p_1,p_2;p_3,p_4)=[2w_a({\bf p}_1) \cdots 2w_d({\bf p}_4)]^{1/2}T_{NR}^{ab;cd}(p_1,p_2;p_3,p_4)\, ,
\en
where the subscripts $R$ and $NR$ label the relativistic and non-relativistic
theories, and $w_a({\bf p})=\sqrt{M_{\pi^a}^2+{\bf p}^2}$. The presence
of the overall factor in the matching condition~(\ref{eq:matching}) reflects 
the difference in the normalization of the one-particle states and the 
field operator in the non-relativistic and relativistic theories.
It is understood that both sides of this equation are expanded in powers
of the momenta ${\bf p}_i$. The matching should therefore be performed at a given
order in the momentum expansion -- it fixes
 the polynomial parts of the amplitudes in all physical channels.
This is exactly the freedom one has in choosing the couplings
of the non-relativistic Lagrangian.
On the other hand, the non-analytic pieces
proportional to $\sigma_c,\sigma_0$ are reproduced  automatically, according 
to analyticity and unitarity, which hold both in the relativistic and in the non-relativistic theories.

This non-relativistic effective theory
obeys  {\em power counting rules} in a generic small 3-momentum $p$: 
 Bubble diagrams with charged (neutral) pions running in the loop are  proportional
to $p_c$ ($p_0$). Consequently,  multi-loop diagrams are 
suppressed by pertinent powers of $p_c,p_0$. It can be  shown that
the relativistic insertions and derivative
couplings in the diagrams do not destroy the power counting.

The diagrammatic expansion in this theory coincides with the {\em
effective range
expansion}.
 This can be seen most easily, if one assumes isospin symmetry $M_\pi=M_{\pi^0}$.
In this case, the bubbles vanish at threshold where $p_c=p_0=0$, and so do
the contributions from the derivative vertices.
This means that, 
to all orders, the threshold amplitudes are determined in terms of
the non-derivative couplings $c_1,c_2,c_3$. Using the matching 
condition~\eqref{eq:matching},
 one
can  express these couplings through the $\pi\pi$ scattering lengths 
with definite isospin,
\eq\label{eq:match_c1c2c3}
3M_\pi^2c_1&=&4\pi(2a_0+a_2)+\cdots\, ,
\nonumber\\[2mm]
3M_\pi^2c_2&=&4\pi(a_2-a_0)+\cdots\, ,
\nonumber\\[2mm]
3M_\pi^2c_3&=&2\pi(a_0+2a_2)+\cdots\, ,
\en
where the ellipses stand for  isospin-breaking corrections.
Analogously, the derivative couplings in the Lagrangian can be expressed 
through effective ranges, shape parameters, etc. This property is ideally
suited for describing hadronic atoms: the scattering lengths, which we 
want to extract from experimental data, turn out to be
 the parameters of the 
Lagrangian which will be used to describe the atoms. Consequently, the 
calculation of atomic observables in perturbation theory by using this 
Lagrangian will automatically generate 
 a parametrization of the former directly in terms of  scattering lengths.

\subsection{Including photons}

The inclusion of  virtual photons in this framework
is straightforward. First, one follows the paradigm of minimal coupling
 and replaces  ordinary space-time derivatives of
the charged pion fields  by  covariant ones. In addition, the Lagrangian
contains the kinetic term for free photons and 
a tower of gauge- and rotationally-invariant operators, which can be
built from the electric ${\bf E}$ and magnetic ${\bf B}$ fields. For example,
the  kinetic term for charged pions becomes
\eq 
\hspace*{-.4cm} 
{\cal L}_{kin}^\pm=\sum_\pm\Phi_\pm^\dagger\biggl(iD_t-M_\pi
+\frac{{\bf D}^2}{2M_\pi}+\frac{{\bf D}^4}{8M_\pi^3}+\cdots\mp
eh_1\frac{{\bf D}{\bf E}-{\bf E}{\bf D}}{6M_\pi^2}+\cdots\biggr)\Phi_\pm\, ,
\en
where $D_t\Phi_\pm=(\partial_t\mp ieA_0)\Phi_\pm,{\bf D}\Phi_\pm={\nabla}\Phi_\pm\pm ie{\bf A}\Phi_\pm$ are  covariant derivatives, $e$ is the electric charge, 
and $(A_0,{\bf A})$ 
 denotes the photon field. Furthermore, $h_1$ 
is a new LEC,  related to the electromagnetic
radius of the pion, $h_1=M_\pi^2\langle r_\pi^2\rangle+O(\alpha)$.
 
The power-counting at tree-level, which amounts to counting the number of 
3-momenta in a given Feynman diagram, can be carried out analogously to
 the case without
photons. However, loop corrections in general 
lead to a breakdown of  naive
power-counting rules. This is a well-known problem, caused by the
presence of a heavy scale $M_\pi$ in the Feynman integrals:
 loop integrals receive contributions from 
regions where the integration momenta are of the order of $M_\pi$,
 which cause a breakdown
of the counting rules. On the other hand, the effect can be completely
removed by simply
changing the renormalization prescription in the non-relativistic EFT. This is so 
 because the terms
which break power counting behave like polynomials at low energy. 
Most straightforwardly, the goal can be achieved by modifying the prescription
for the evaluation of  Feynman integrals. The pertinent modification is
called  ``threshold expansion''~\cite{Beneke1,Beneke2}. 
A detailed description in the context of the hadronic atom problem can be
found, e.g., in Refs.~\cite{Bern4,Physrep}. In brief, the method
boils down to Taylor-expanding the integrand in any Feynman integral in powers
of the 3-momenta prior to performing the loop integrals in  dimensional
regularization. The expansion and the integration do not commute: it can be
shown that the two results differ by
just the above-mentioned polynomial contribution, which is
absent in the threshold-expanded integral. Thus, applying threshold expansions
to all loop integrals leads to a restoration of the naive power counting
rules in the non-relativistic EFT.

\begin{figure}[t!]
\begin{center}
\includegraphics[width=13.5cm]{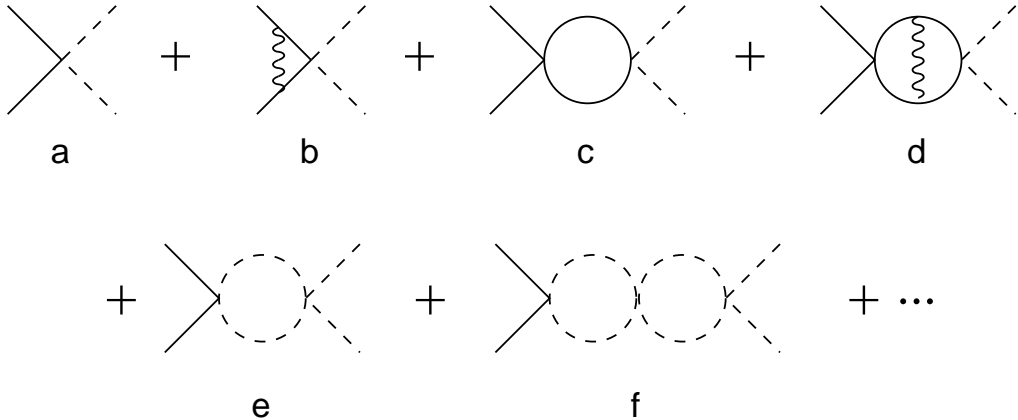}
\end{center}
\caption{\baselineskip 12pt 
Representative set of diagrams contributing to the scattering
  amplitude of the process $\pi^+\pi^-\to\pi^0\pi^0$ at the order of accuracy
  we are working. The solid and dashed lines denote charged and neutral pions,
respectively, and the wiggled line denotes the Coulomb photon (the
calculations are done in the Coulomb gauge). Transverse photons do not
contribute at this order. Only the diagrams a,d,f contribute
to the expression of the real part of the threshold amplitude, see
Eq.~(\ref{eq:TTT}).
}
\label{figure4}
\end{figure}

In principle, the matching condition in the presence of photons is again given by
the relation Eq.~(\ref{eq:matching}). On the other hand, the
scattering amplitudes in the presence of real and virtual  photons 
 are infrared-divergent  in
perturbation theory. It is then natural to identify
non-singular parts of the amplitude,  which are more convenient for matching.
At the accuracy needed here, it suffices to discuss
the problem at order $e^2$, for the charge-exchange process
$\pi^+\pi^-\to\pi^0\pi^0$. The structure of the scattering amplitude 
in the vicinity of the threshold $|{\bf p}|\to 0$ is identical in the relativistic
and in the non-relativistic case,
\eq\label{eq:curlyA}
{\rm e}^{-i\alpha\theta_c}T^{+-;00}=\frac{e^2b_1}{|{\bf p}|}
+e^2b_2\ln\frac{2|{\bf p}|}{M_\pi}+{\cal T}^{+-;00}+O({\bf p})\, ,
\en
where ${\bf p}$ denotes the relative 3-momentum in the CM frame and
$\theta_c$ is the (infrared-divergent) Coulomb phase,
\eq
\theta_c=\frac{M_\pi}{2|{\bf p}|}\,\mu^{d-3}
\biggl\{\frac{1}{d-3}-\frac{1}{2}\,[\Gamma'(1)+\ln 4\pi]
+\ln\frac{2|{\bf p}|}{\mu}\biggr\}\, .
\en 
The scale $\mu$ is generated by  dimensional regularization, which
 is used to tame  infrared and ultraviolet
divergences.
 The coefficients $b_{1,2}$ differ by a factor $4M_\pi^2$ in the relativistic and
 in the non-relativistic theory. 
Finally,  ${\cal T}^{+-;00}$ 
denotes the {\em threshold amplitude}, which is the counterpart of the
scattering length in the presence of photons. It is infrared-finite.
The matching condition at threshold  reads 
\eq\label{eq:matching_gamma}
{\cal T}_R^{+-;00}=4M_\pi^2\,{\cal T}_{NR}^{+-;00}\, .
\en
Calculating the diagrams of the type shown in Fig.~\ref{figure4}, we arrive
at an expression for ${\cal T}_{NR}^{+-;00}$ in terms of the 
non-relativistic couplings $c_1,c_2,c_3,\cdots$. At the order of accuracy
we are working, only a finite number of diagrams contribute. 
The final result for the { real part} of the
threshold amplitude is given by
\eq\label{eq:TTT}
\mbox{Re}\,{\cal T}^{+-;00}_{NR}&=&2c_2-c_2c_3^2\,\frac{\Delta_\pi M_{\pi^0}^2}
{2\pi^2}
+c_1c_2\frac{\alpha M_\pi^2}{4\pi}\biggl(1-\Lambda(\mu)
-\ln\frac{M_\pi^2}{\mu^2}\biggr)+o(\delta)\, ,
\nonumber\\
\en
where $\Delta_\pi=M_\pi^2-M_{\pi^0}^2$, and $\Lambda(\mu)$ stands for the
{\em ultraviolet} divergence originating from the diagram in
Fig.~\ref{figure4}d,
\eq
\Lambda(\mu)=\mu^{2(d-3)}\biggl\{\frac{1}{d-3}-\Gamma'(1)-\ln 4\pi \biggr\}\, .
\en 
The ultraviolet divergence is removed in a standard manner, by renormalizing
the coupling $c_2$.

The matching condition 
(\ref{eq:matching_gamma}) enables one to relate a particular combination of
the couplings to the relativistic threshold amplitude ${\cal T}_R^{+-;00}$. 
At the accuracy needed here, higher-order terms in the momentum expansion
 of the amplitudes are not needed.

Finally, we note that the non-relativistic couplings $c_i$ contain both, strong and
electromagnetic isospin-breaking
corrections. According to the unified 
counting of the isospin-breaking effects, which
was introduced in section~\ref{sec:essentials}, we write
\eq
c_i=\bar c_i+\alpha c_i^{(1)}+(m_d-m_u)^2c_i^{(2)}+o(\delta)\, ,
\en
where the bar denotes quantities taken in the isospin limit 
$\alpha=0,~m_d=m_u$. The $\bar c_i$ can be related to scattering lengths and effective ranges in the isospin symmetric world. On the other hand, 
the coefficients  $c_i^{(1)}, c_i^{(2)}$ are fixed
via matching to ChPT.

\subsection{Bound states}

The non-relati\-vis\-tic framework does not contain any new dynamical
information about the behavior of the scattering amplitudes at low momenta,
because it is constructed  such that it reproduces the relativistic amplitudes.
However, the non-relativistic framework is extremely useful  
when  bound states are considered, because methods of 
standard quantum mechanics can be used to a large extent.
As all couplings in the non-relativistic Lagrangian 
have been fixed through  matching of the scattering amplitudes, there are 
no additional free 
parameters left in the bound-state sector. Consequently, 
solving the bound-state problem in the non-relativistic theory,
 one can eventually express
the observables of the bound states in terms of the parameters of the
{\em relativistic} scattering amplitudes. We now describe the procedure.

Hadronic atoms are shallow quasi-stable states
formed predominately by the Coulomb force. In order to describe such states, 
it is convenient to use perturbation theory, where the unperturbed solution
corresponds to the purely Coulombic bound state.
The full Hamiltonian of the system is constructed from the Lagrangian with
standard methods. Here, we concentrate on that part of the Hamiltonian 
which is responsible for the next-to-leading-order term in the 
DGBT formula. Moreover, we
confine for simplicity the calculation  to the {\it width} of the ground state. 
 As shown in Ref.~\cite{Bern4}, the pertinent Hamiltonian is
\eq\label{eq:H}
{\bf H}={\bf H}_0+{\bf H}_C+{\bf H}_S+{\bf H}_R+\cdots=
{\bf H}_0+{\bf  H}_C+{\bf V}+\cdots\, ,
\en
where ${\bf H}_{0,C,S,R}$ stand for the free non-relativistic Hamiltonian,
Coulomb interaction, strong interactions and the relativistic
corrections to the pion kinetic energy.
 [It is convenient to use Coulomb gauge in the
  non-relativistic calculations. In this gauge, transverse photons can be
  dropped completely, since they do not contribute at this accuracy to the width. The
  time-like photon field can be eliminated by using equations of motion,
  resulting into the static Coulomb potential acting between pions. Finally
  note that it is legitimate to use different gauges in the relativistic and
  non-relativistic theories, since only gauge-invariant quantities enter the
  matching condition.] The ellipses in Eq.~(\ref{eq:H}) stand for  terms that 
 do not contribute to the width
  of the ground state at next-to-leading order.
  The explicit expressions read
\eq\label{eq:HHHH}
{\bf H}_0&=&\int d^3{\bf x}\,\sum_{a=\pm,0}
\Phi_a^\dagger({\bf x},0)
\biggl(M_{\pi^a}-\frac{\triangle}{2M_{\pi^a}}\biggr)
\Phi_a({\bf x},0) 
\nonumber\\[2mm]
{\bf H}_C&=&-\frac{e^2}{4\pi}\int d^3{\bf x}d^3{\bf y}\,
(\Phi_-^\dagger({\bf x},0)\Phi_-({\bf x},0))\,\frac{1}{|{\bf x}-{\bf y}|}\,
 (\Phi_+^\dagger({\bf y},0)\Phi_+({\bf y},0))\, ,
\nonumber\\[2mm]
{\bf H}_S&=&\int d^3{\bf x}\,\biggl\{
-c_1\,\Phi_+^\dagger\Phi_-^\dagger\Phi_+\Phi_-
-c_2\,(\Phi_+^\dagger\Phi_-^\dagger\Phi_0^2+\mbox{h.c.})
-c_3\,(\Phi_0^\dagger\Phi_0)^2\biggr\}\, ,
\nonumber\\[2mm]
{\bf H}_R&=&\int d^3{\bf x}\,\sum_{a=\pm,0}
\Phi_a^\dagger({\bf x},0)\biggl(-\frac{\triangle^2}{8M_{\pi^a}^3}\biggr)
\Phi_a({\bf x},0)\, . 
\en
  It is seen that the Hamiltonian is 
amazingly simple: it contains three couplings $c_{1,2,3}$ that need to be 
matched -- all the rest is known.

The pure Coulomb state is an eigenstate
of the Hamiltonian ${\bf H}_0+{\bf H}_C$.
 The resolvent 
${\bf G}_C(z)=(z-{\bf H}_0-{\bf H}_C)^{-1}$
develops a tower of  poles on the negative real axis in the complex
 $z$- plane, 
at $z=E_n=2M_\pi-\alpha^2M_\pi/4n^2,~n=1,2,\cdots$ (in the CM frame).
The position of these poles coincide with the Coulomb binding energies.
 Once  the perturbation
${\bf V}$ is switched on, the poles move from the real axis 
to the second Riemann sheet in
the complex $z$-plane. The energy shift and width of a given state
is defined by the real and imaginary parts of the shifted pole position.
Restricting ourselves to the ground state, we write
\eq
\Delta E-i\,\frac{\Gamma}{2}=z-E_1\, ,
\en
The shift $z-E_1$ of the pole position  can be consistently treated
with the Feshbach formalism~\cite{Feshbach1,Feshbach2}. 
A detailed discussion thereof in the context of hadronic atoms can be found in 
Refs.~\cite{Bern1,Bern4,Physrep}. There, it is shown that the shift
is given by the standard 
expression known from the Rayleigh-Schr\"odinger perturbation theory,
\eq\label{eq:key}
z-E_1=\langle\Psi_G|\,\biggl\{{\bf V}+{\bf V}\,\sum_{E_\alpha\neq E_1}
\frac{|\Psi_\alpha\rangle\langle\Psi_\alpha|}{z-E_\alpha}\,{\bf V}
+\cdots\biggr\}\,|\Psi_G\rangle\, ,
\en
where the sum over $\alpha$ runs over both the discrete and
continuous spectra of the unperturbed Hamiltonian ${\bf H}_0+{\bf H}_C$,
and $|\Psi_G\rangle$ denotes the ground-state vector. 
 In momentum space,
\eq\label{eq:Psiground}
\Psi_G({\bf k})=\frac{(64\pi\gamma^5)^{1/2}}{({\bf k}^2+\gamma^2)^2} \, ,
\quad\quad \gamma=1/r_B=\alpha M_\pi/2\, .
\en 
The center-of-mass (CM) 
 motion is removed in the above matrix elements,
which are then evaluated in the CM frame ${\bf P}=0$. 

It is instructive to first neglect the relativistic corrections ${\bf H}_R$.
At  leading order in ${\bf V}$, the shift of the pole position is real. Using
Eqs.~(\ref{eq:HHHH}), (\ref{eq:key}) and (\ref{eq:Psiground}), we get
\eq\label{eq:OV}
\Delta E=-\frac{\alpha^3 M_\pi^3}{8\pi}\,c_1+O({\bf V}^2)\, ,\quad\quad
\Gamma=O({\bf V}^2)\, .
\en
The matching condition displayed in Equation (\ref{eq:match_c1c2c3})
 finally leads to
\eq
\Delta E=-\frac{1}{6}\,\alpha^3M_\pi\,(2a_0+a_2)+\cdots\fs
\en
The ellipses denote contributions of order $\delta^4$.

The decay width is of  order ${\bf V}^2$. This leading term is generated by the
contribution from  the two neutral pion intermediate state in Eq. \eqref{eq:key}.
 The pertinent threshold is below the bound state energy - these states 
therefore generate an imaginary part in the energy shift. 
Since neutral pions do not feel the Coulomb
potential, the sum over those intermediate states in Eq.~(\ref{eq:key})
merely yields the bubble integral with two neutral pions, similar to the one  
displayed in Eq.~(\ref{eq:Jbubble}). The  result for the width at this order reads
\eq
\Gamma=-2\,\mbox{Im}\,z\scs
\en
where $z$ is a solution to the equation
\eq\label{eq:solutionz}
z=-\frac{\alpha^3M_\pi^3}{4\pi}\,c_2^2\,J_0(z)\fs
\en
Here, $J_0(z)$ is given in  Eq.~(\ref{eq:Jbubble}), with
$M_\pi$ replaced by $M_{\pi^0}$. The equation \eqref{eq:solutionz}
has a solution on the second Riemann sheet only. The width becomes
\eq\label{eq:Gamma}
\Gamma=\frac{\alpha^3M_\pi^3M_{\pi^0}}{8\pi^2}\,\rho^{1/2}c_2^2+\cdots
=\frac{2}{9}\,\alpha^3\rho^{1/2}(a_0-a_2)^2+\cdots\, ,
\en
where $\rho=2M_{\pi^0}(M_\pi-M_{\pi^0}-M_\pi\alpha^2/8)$.  In the last step,
the matching condition (\ref{eq:match_c1c2c3}) was used. 

It is seen that the result for the width is of order $\delta^{7/2}$ at leading order. 
To work out the next-to-leading order terms,
 one has to include in  Eq.~(\ref{eq:key}) contributions up 
to and including $ {\bf V}^3$, with ${\bf V}={\bf H}_S+{\bf H}_R$. 
According to  power counting, the subsequent terms  are
suppressed by  positive 
powers of $\delta$. This can be seen e.g. from
Eqs.~(\ref{eq:Jbubble}) and (\ref{eq:pcp0}), showing that the charged
and neutral bubbles, evaluated at the bound-state energy $P^0=E_1$, count as
$O(\delta)$ and $O(\delta^{1/2})$, respectively. This is a very important
property of the non-relativistic EFT in dimensional regularization:
 at a given order in $\delta$, only a finite number of terms in the perturbation series
contribute.

Finally, the result for the decay width of pionium
up to and including terms of order $\delta^{9/2}$ reads
\eq\label{eq:Gamma_full}
\Gamma=\frac{\alpha^3M_\pi^3M_{\pi^0}}{8\pi^2}\,\rho^{1/2}\,c_2^2
\biggl(1+\frac{5\rho}{8M_{\pi^0}^2}\biggr)
\biggl(1-\frac{\rho M_{\pi^0}^2c_3^2}{4\pi^2}\biggr)
(1-2c_1g(E_1))\, ,
\en
where $g(E_1)$ corresponds to the sum of diagrams
where any number of Coulomb photons is exchanged between the charged 
pions. The explicit expression for this quantity is
given by
\eq\label{eq:Schwinger}
g(E_1)=\frac{\alpha M_\pi^2}{8\pi}
\biggl(2\ln\alpha-3+\Lambda(\mu)+\ln\frac{M_\pi^2}{\mu^2}\biggr)\, .
\en

\noindent
Using the matching condition (\ref{eq:TTT}), one may finally express the decay
width through the relativistic threshold amplitude of the process
$\pi^+\pi^-\to\pi^0\pi^0$,
\eq\label{eq:final}
\Gamma&=&\frac{2\alpha^3p^\star}{(32\pi)^2}\,(\mbox{Re}\,{\cal T}_R^{+-;00})^2
(1+K)+o(\delta^{9/2})\, ,
\nonumber\\
K&=&\frac{\Delta_\pi}{9M_\pi^2}\,(a_0+2a_2)^2
-\frac{2\alpha}{3}\,(\ln\alpha-1)\,(2a_0+a_2)\, .
\en
We note that the reference to the non-relativistic theory has completely 
disappeared in the final result Eq.~(\ref{eq:final}): the decay width is expressed
through the {\em relativistic} threshold amplitude.

We have thus achieved the first {\it main goal } mentioned at the end of 
section \ref{sec:essentials}.
 It remains to express the relativistic threshold amplitude  in terms 
of isospin symmetric scattering lengths. Then, one can extract these 
from the measured lifetime of the pionium ground state.

\subsection{Scattering lengths}
\label{subsec:ChPT}

The prediction for the scattering lengths $a_{0,2}$ in Eq.~\eqref{eq:a0a2}
 concerns an isospin-symmetric 
 {\em paradise world} --   QCD at $m_u=m_d$. 
In this world, there are no electromagnetic interactions.
The light quark masses $m_u=m_d, m_s$ and the scale $\Lambda_{QCD}$ are chosen such
that $M_\pi=M_{\pi^+}=139.57$ MeV, $M_K=493.68$ MeV, $F_\pi$=92.4 MeV. The precise values of the heavy quark masses $m_{c,b,t}$ do not matter in the present context.
 On the other hand, 
the threshold amplitude, which occurs in 
the DGBT formula, concerns the real world, 
where $m_u\neq m_d,\alpha\neq 0$. We are thus faced with the problem to relate
 that amplitude  to the scattering lengths evaluated in the paradise world.

The  structure of the threshold amplitude at $\alpha\neq 0, m_u\neq m_d$ is
\eq\label{eq:structure}
-\frac{3}{32\pi}\,{\mbox{Re} \, \cal T}_R^{+-;00}=a_0-a_2+h_1(m_d-m_u)^2+h_2\alpha+o(\delta)\, ,
\en
where the coefficients $h_i$ can be systematically 
calculated in the framework of ChPT. 
These calculations  are carried out for the scattering amplitude, not for
the bound state observables.
 Thus, the use of the non-relativistic approach enables
one to separate bound state calculations from the chiral expansion.
  
We  outline the determination of $h_{1,2}$ at order $p^2$. 
The leading order Lagrangian of ChPT is 
\eq
{\cal L}_2=\frac{F^2}{4}\,\langle \partial_\mu U\partial^\mu U^\dagger
+2B{\cal M}(U+U^\dagger)\rangle+C\langle QUQU^\dagger\rangle\, ,
\en
where the unitary matrix $U$ contains the pion fields, 
$\langle\cdots\rangle$ denotes the trace in flavor space, and
\eq
{\cal M}=\mbox{diag}\,(m_u,m_d)\, ,\quad\quad
Q=\frac{e}{3}\,\mbox{diag}\,(2,-1)
\en
are the quark mass matrix and the charge matrix, respectively. Finally, the
constant $C$ is related to the charged and neutral pion mass difference
$M_\pi^2-M_{\pi^0}^2= 2e^2C/F^2$.

The threshold scattering amplitude $\pi^+\pi^-\to \pi^0\pi^0$ at order $p^2$ is given by
\eq
{\cal T}_R^{+-;00}=-\frac{s-M_{\pi^0}^2}{F^2}\biggr|_{s=4M_\pi^2}
=-\frac{3M_\pi^2}{F^2}-\frac{\Delta_\pi}{F^2}\, .
\en
From this result, the expressions for $h_i$ at leading order can be
 read off,
\eq\label{eq:leading}
h_1=O(\hat m)\, ,\quad\quad
h_2=\frac{3\Delta_\pi}{32\alpha\pi F^2}+O(\hat m)\, .
\en
The details of the calculation at  next-to-leading order can be found, e.g.,
in Refs.~\cite{Bern2,Bern4}. The result for the width at next-to-leading order is
\eq
\Gamma=\frac{2}{9}\alpha^3p^\star(a_0-a_2)^2(1+\delta_\Gamma)\, ,\quad\quad   
\delta_\Gamma=(5.8\pm 1.2)\times 10^{-2}\, .
\en
\begin{sloppypar}
\noindent
Note that the bulk of the total correction is generated by the leading-order
term~(\ref{eq:leading}), which contains no free parameters. We expect
that next-to-next-to-leading corrections will be completely negligible.
Vacuum polarization has been investigated in pionium and/or other atoms in References~\cite{Karshenboim,Labelle1,Eiras,Bern4,Physrep}. 
\end{sloppypar}

Using the scattering lengths  in Eq.~(\ref{eq:a0a2}), we  arrive
at the prediction for the pionium lifetime~\cite{Bern4}, 
\eq\label{eq:tau}
\tau=\frac{1}{\Gamma}=(2.9\pm 0.1)\times 10^{-15}\,\mbox{s}\, .
\en
The result of the ongoing measurement carried out by the DIRAC 
collaboration  agrees with this  value,
\eq\label{eq:tauDIRAC}
\tau=\left(2.91^{+0.49}_{-0.62}\right)\times 10^{-15}\,\mbox{s}
\quad [\mbox{DIRAC, Ref.~\cite{DIRAC-2.9}}]\fs
\en
It is expected that the precision of the measurement improves in 
the near future, see Ref.~\cite{Zhabitsky}.

\setcounter{equation}{0}
\section{Pionic hydrogen and
  pionic deuterium}

\label{sec:piH}

The power and beauty of the non-relativistic effective Lagrangian approach is
best demonstrated by the fact that the description of {\em all} hadronic atoms,
which were mentioned in the introduction, proceeds very similarly to the pionium 
case just discussed.
 On the other hand, some of these bound 
systems are very different {\em physically}, and so are the results obtained.
The important point is that
 the language used to describe these systems stays  -- with only
minor modifications -- always the same.

As an example, we consider in this section the measurement of the $S$-wave $\pi N$
scattering lengths $a_{0+}^+,~a_{0+}^-$ in  experiments on pionic hydrogen and pionic
deuterium, which are performed by Pionic Hydrogen collaboration at
PSI~\cite{PSI1,PSI2,PSI3,PSI4,PSI5}. 
Measuring the energy shift and the width  enables one to extract 
very accurate values of the real and
imaginary parts of the elastic $\pi^-p$ threshold scattering amplitude, using
pretty much the same technique as in the pionium case.
Using unitarity and the measured Panofsky ratio  finally allows one to
 to extract from data the real part of the
threshold amplitudes for both, the elastic $\pi^-p\to \pi^-p$ and the
charge-exchange $\pi^-p\to \pi^0n$ reactions.

In the last step, the threshold amplitudes are again  related 
to the pertinent scattering lengths in the isospin-symmetric world (cf. with subsection~\ref{subsec:ChPT}).
At  leading order, the relation is
\eq\label{eq:iso-piN}
{\cal T}^{\pi^-p\to\pi^-p}&=&a_{0+}^++a_{0+}^-
+\frac{1}{4\pi (1+M_\pi/m_p)}\,\biggl(\frac{4\Delta_\pi}{F_\pi^2}\, c_1
-\frac{e^2}{2}\, (4f_1+f_2)\biggr)\, ,
\nonumber\\[2mm]
{\cal T}^{\pi^-p\to\pi^0n}&=&-a_{0+}^-+\frac{1}{16\pi (1+M_\pi/m_p)}\,
\biggl(\frac{g_A^2\Delta_\pi}{m_pF_\pi^2}+2e^2f_2\biggr)\, ,
\en
where $m_p$ denotes the nucleon mass, $g_A$ is the axial coupling constant of
the nucleon and $c_1,f_1,f_2$ are various (strong and electromagnetic) LECs
from the second-order pion-nucleon 
Lagrangian~\cite{Mojzis,pin-lagr1,pin-lagr2}. 
(Following the tradition in the literature, 
we use the same notation for the LEC $c_1$ as in pionium. 
We hope that this is not confusing.)

The difference between  pionium and  pionic hydrogen becomes
visible by comparing Eqs.~(\ref{eq:structure},\ref{eq:leading}) and (\ref{eq:iso-piN}): 
whereas the pertinent isospin-correction for
 pionium at leading order is parameter-free,  Eq.~(\ref{eq:iso-piN}) contains 
the LECs $c_1,f_1,f_2$, whose values are
not established very well. The issue has been discussed in detail in
Ref.~\cite{Physrep} where, in particular, an update on the values of 
$c_1,f_2$ can be found.
No reliable determination on the basis of  experimental input is
available for $f_1$  at present. This is the reason for a
substantial uncertainty in  the leading correction in the $\pi N$ case,
 which by far exceeds the
experimental error in the measurement of the energy shift. 
 
Next, we turn to pionic deuterium, which  allows one to
extract the pion-deuteron threshold scattering amplitude. However, in this
case the analysis is not yet complete: what one intends to finally obtain are the
pion-nucleon scattering lengths, which are related to the pion-deuteron
amplitude through   multiple-scattering theory. This is a very complicated
issue, which has been extensively addressed in the past within the framework
of  potential models. Recently,  calculations in  EFT have been  
performed as well
 (see, e.g., Refs.~\cite{Raha2,Raha3,Weinberg-deuteron,Bernard,
Hanhart_Ko1,Hanhart_Ko2,Bernard2_1,Bernard2_2,Bernard2_3,%
Valderrama-Platter1,Valderrama-Platter2,Doring,Epelbaum,%
Borasoy-Beane1,Borasoy-Beane2}).
This method  allows one to largely reduce an uncontrolled systematic error in the 
resulting values of the $\pi N$ scattering lengths.

The calculations within EFT have shed new light on
the importance of isospin-breaking corrections, a point which is 
obscure in 
potential models. Namely, the pion-deuteron scattering length in the
isospin limit vanishes at leading  order. For this reason, the
isospin-breaking correction to this quantity, determined pre\-domi\-na\-tely by
  {\em short-range} physics, turns out to be very large~\cite{Raha3}.
In the context of the pion-nucleon scattering, the same effect has been
mentioned in Ref.~\cite{Weinberg-isobr}. Further, the isospin-breaking
correction to the pion-deuteron threshold amplitude contains
the same virtually unknown LEC $f_1$ as the $\pi^-p$ elastic scattering
amplitude and is therefore determined with a large systematic error.

The experiments on  pionic hydrogen and pionic deuterium are complementary
to each other. Namely, the data on pionic hydrogen alone determine the scattering
lengths $a_{0+}^+$ and $a_{0+}^-$ separately. The data on the energy shift of
pionic deuterium provides an additional constraint on these two quantities. This 
can be seen by considering the bands in the $(a_{0+}^+,a_{0+}^-)$
plane, which correspond to the different observables. Measuring each of the
following three observables: the energy shift and width of pionic hydrogen and
the energy shift of  pionic deuterium  fix a particular combination of
$a_{0+}^+$ and $a_{0+}^-$. Each combination
 corresponds to a band, whose width is
determined by a combined experimental and theoretical error. If these three
bands do not have a common intersection area in the $(a_{0+}^+,a_{0+}^-)$ 
plane, then either experiment or/and the
theoretical interpretation of the data is not correct. 
\begin{figure}[t!]

\vspace*{.8cm}
\begin{center}
\includegraphics[width=11.cm]{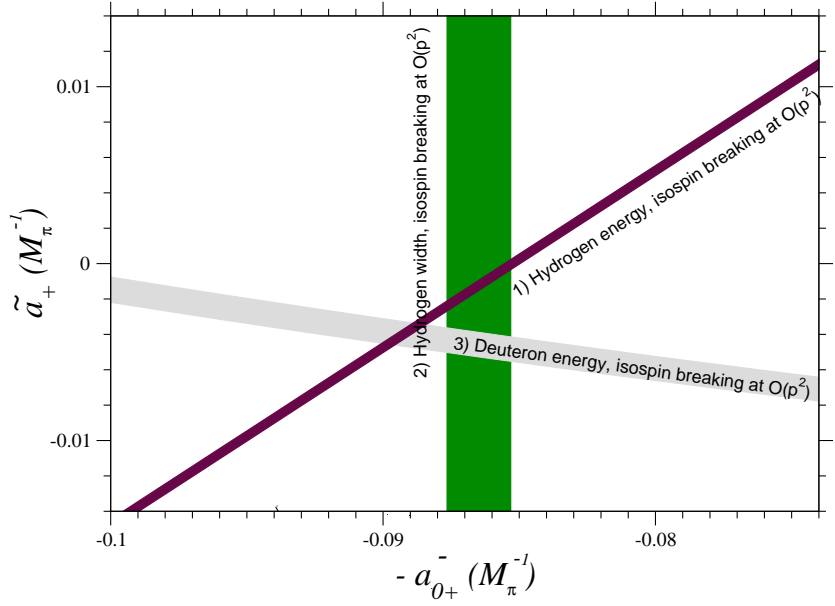}
\end{center}
  \caption{\baselineskip 12pt 
The quantity $\tilde a^+$ defined in Eq.~(\ref{eq:atilde}),
 plotted against $a_{0+}^-$. Three different bands emerge from the measurement 
of different observables. It is seen that with the isospin breaking
    corrections evaluated at $O(p^2)$, the three bands 1), 2) and 3) 
have no common intercept.  }
\label{figure5}
\end{figure}

The consistency check is made 
complicated by a large uncertainty present in the values of LECs -- most notably in
$c_1$ and $f_1$.  Baru et al. in Ref.~\cite{barumenu2007} have proposed a procedure to
partially circumvent the problem. The idea is based on the observation that
the LECs $c_1$ and $f_1$ enter in the same combination 
in both, the pionic hydrogen and pionic deuterium energy shift (at leading chiral
order). Introducing the quantity
\eq\label{eq:atilde}
\tilde a_+=a_{0+}^++\frac{1}{4\pi(1+M_\pi/m_p)}\,
\biggl(\frac{4\Delta_\pi}{F_\pi^2}\, c_1-2e^2f_1\biggr)\, ,
\en
it is seen that $c_1$ and $f_1$ disappear at leading order 
from the expressions of the
hadronic atom observables, if these are written  in terms of $\tilde a_+$ and $a_{0+}^-$.
Hence, various bands in the plot shown in Fig.~\ref{figure5} are much
narrower than the pertinent bands that can be 
drawn in  $(a_{0+}^+,a_{0+}^-)$ plane (see Ref.~\cite{Physrep}).
In particular, Fig.~\ref{figure5} demonstrates that three bands still fail
to pass this elaborate consistency test. However, it can be argued that
next-order isospin-breaking corrections can be large and may change the above
picture. In order to carry out a meaningful test, these corrections should be
calculated at least to $O(p^3)$ for all three observables.
The possibility to perform the consistency check at  higher 
accuracy, however, is not yet the end of the story. As it can be seen from
Eq.~(\ref{eq:atilde}), the relation of the isospin-symmetric scattering
length $a_{0+}^+$ to the quantity $\tilde a_+$ does contain both, $c_1$ and
$f_1$. Thus, in order to determine  $a_{0+}^+$ at a reasonable accuracy, one
should find ways to estimate these LECs at the required precision.

Finally, we mention that the extraction of the $\bar KN$ 
scattering lengths from the experimental data on kaonic hydrogen and kaonic
deuterium~\cite{SIDDHARTA1,SIDDHARTA2,SIDDHARTA3,SIDDHARTA4,KEK1,KEK2} 
bears many similarities to the pion-nucleon
case. The analysis of the problem within three-flavor ChPT, 
however, 
is more complicated due to the large value of the strange quark mass. In 
addition, the presence of the sub-threshold $\Lambda(1405)$ re\-so\-nan\-ce leads to
a large $S$-wave scattering length. As a result, the deuteron problem can no
more be treated purely perturbatively, and a partial re-summation of the
multiple-scattering series should be considered. These very interesting
issues, however, cannot be covered in the present review. The interested
reader is referred to the original publications, e.g., Refs.~\cite{Raha1,Raha4,%
Dalitz1,Dalitz2,Dalitz3,Kamalov}.  
 
\setcounter{equation}{0}
\section{Summary points and future issues}
\label{sec:concl}

\begin{itemize}

\item[i)]
Precise data on the energy levels and lifetimes of hadronic atoms enable 
one to extract various hadronic scattering lengths, provided that
a systematic method to work out the relation 
between data and scattering lengths is available.
\item[ii)]
As we discussed in this review, a very convenient framework  is provided
by the non-relativistic effective Lagrangian approach.
 Its non-relativistic feature are used in intermediate steps only -- 
 at the end of the calculations, all observables are expressed in terms 
of the underlying relativistic theory, QCD+QED.
\item[iii)]
Despite the fact that various hadronic atoms observed in Nature are governed
by  very different underlying physics, the same framework based on the 
non-relativistic effective Lagrangians applies -- with minor modifications -- 
to all of them. This is a beautiful demonstration of the potential and the 
flexibility of  non-relativistic EFT.
\item[iv)]
To date, the conceptual problems of the general theory of hadronic atoms have been 
clarified to a large extent. Now, the focus shifts mainly to  applications.
Among these, we mention the evaluation of a full set of isospin-breaking 
corrections at third order in pionic hydrogen and in pionic deuterium.  
In addition, it would be a major breakthrough to present a systematic 
calculation of the kaon-deuteron scattering length in terms of the threshold
parameters of the $\bar KN$ interaction beyond the static approximation.

\end{itemize}

\section*{Disclosure statement}
The authors are not aware of any biases that might be perceived 
as affecting the objectivity of this review.

\bigskip

\noindent{\em Acknowledgments:}

\bigskip

\begin{sloppypar}
 We thank Ulf-G. Mei\ss ner for useful comments which helped us to improve  the 
manuscript. The Center for Research and Education in Fundamental Physics is
  supported by the ``Innovations- und Kooperationsprojekt C-13'' of
  the ``Schweizerische Universit\"atskonferenz SUK/CRUS''.
This work was  supported  by the Swiss
National Science Foundation,  and by EU MRTN-CT-2006-035482
(FLAVIA{\it net}). 
  Partial financial support under the EU Integrated Infrastructure
Initiative Hadron Physics Project (contract number RII3-CT-2004-506078),
by DFG (SFB/TR 16, ``Subnuclear Structure of Matter''), 
by DFG (FA67/31-1, FA67/31-2, GRK683), the President Grant 
of Russia ``Scientific Schools''  No. 817.2008.2 
and by the Helmholtz Association through funds provided to the 
virtual institute 
``Spin and strong QCD'' (VH-VI-231) is gratefully acknowledged. 
 One of us (J.G.) is grateful to the Alexander von Humboldt-Stiftung and to 
the Helmholtz-Gemeinschaft for the award of  a prize 
 that allowed him to stay at the HISKP at the University of Bonn, 
where part of this work was performed. 
He also thanks the HISKP for the warm hospitality during these stays.

\end{sloppypar}

\bibliography{references-new}

\end{document}